\documentclass[lettersize,journal,twocolumn]{article}
\usepackage{cite}
\usepackage{amsmath,amssymb,amsfonts}
\usepackage{algorithmic}
\usepackage{graphicx}
\usepackage{bm}
\usepackage{subfig}
\usepackage{caption}
\usepackage{balance}
\usepackage{mathtools}
\usepackage{booktabs}
\usepackage{float}
\usepackage{changepage}
\usepackage[left=1.60cm,right=1.60cm,top=1.60cm,bottom=2.40cm]{geometry}
\usepackage{rotating}
\graphicspath {{figures/}}
\usepackage{wrapfig}
\usepackage{placeins}
 \usepackage{hyperref}
\hypersetup{
    colorlinks=true,
    citecolor=blue     
    }
\usepackage{multicol, blindtext}
\usepackage[table]{xcolor}
\usepackage{pgfplots}
 \usepackage{wrapfig}
 \usepackage{xcolor}
 \usepackage{tabularx}
 \usepackage{multirow}
\usepackage{bigstrut}
\usepackage{authblk}
\usepackage[pdftex]{lscape}
\usepackage{rotating}
\usepackage{multicol}

\usepackage{bm,array}
\usepackage{color, colortbl}
\definecolor{Gray}{gray}{0.9}
\definecolor{Grayy}{gray}{0.7}
\usepackage{makecell}
\usepackage{nicematrix}
\usepackage{textcomp}
\providecommand{\keywords}[1]{\textbf{\textit{Keywords---}} #1}

\begin{document}
\title{A Novel Data-Driven Method for the Analysis and Reconstruction of Cardiac Cine MRI}
\author[1]{N. Groun}
\author[2]{M. Villalba-Orero}
\author[2]{E. Lara-Pezzi}
\author[3]{E. Valero}
\author[3]{J. Garicano-Mena}
\author[3]{S. Le Clainche}
\affil[1]{ETSI Aeronáutica y del Espacio and ETSI Telecomunicación - Universidad Politécnica de Madrid, 28040 Madrid, Spain}
\affil[2]{Centro Nacional de Investigaciones Cardiovasculares (CNIC), C. de Melchor Fernández Almagro, 3, 28029 Madrid, Spain}
\affil[3]{ETSI Aeronáutica y del Espacio - Universidad Politécnica de Madrid, 28040 Madrid, Spain and Center for Computational Simulation (CCS), 28660 Boadilla del Monte, Spain}
\date{}                     
\setcounter{Maxaffil}{0}
\renewcommand\Affilfont{\itshape\small}

\maketitle

\begin{abstract}
Cardiac cine magnetic resonance imaging (MRI) can be considered the optimal criterion for measuring cardiac function. This imaging technique can provide us with detailed information about cardiac structure, tissue composition and even blood flow. This work considers the application of the higher order dynamic mode decomposition (HODMD) method to  a set of MR images of a heart, with the ultimate goal of  identifying the main patterns and frequencies driving the heart dynamics. A novel algorithm based on singular value decomposition combined with HODMD is introduced, providing a three-dimensional reconstruction of the heart. This algorithm is applied  (i) to reconstruct corrupted or missing images, and (ii) to build a reduced order model of the heart dynamics.
\end{abstract}


\section{Introduction}\label{sec_introduction}
In cardiology, cine magnetic resonance  images (MRI) are a set of MR images acquired over a number of cardiac cycles. Cine MRI is able to accurately capture the cardiac movement and to assess the flow and function in the heart and vessels.\\
Cine MRI provides parallel cross-sectional images of any organ and it even allows three-dimensional (3D) reconstruction of some human organs, which is extremely useful for studying several organs and diagnosing different diseases.\\
Many methods have been introduced to achieve 3D reconstructions of the heart. Kuwahara and Eiho \cite{kuwahara19913} used gated MRI method, which was able to provide several sets of cross-sectional images of a left ventricle and the whole heart during a cardiac cycle. They were able to reconstruct a 3D image of the left ventricle using three pairs of magnetic resonance (MR) images, each pair  taken across a different axis. Salustri and Roelandt \cite{salustri1995three} used their own image acquisition technique to achieve the 3D reconstruction of the heart. After developing a rotational imaging probe in 1994 (\cite{roelandt1994precordial,roelandt1994ultrasonic}), they introduced their basic steps to reconstruct the heart: after the acquisition of the cross-sectional images, (i)  the images were re-sample and converted from polar to cartesian coordinates, (ii) an interpolation step was achieved to fill the space between individual cross sections and finally, (ii) enhanced and displayed the three dimensional reconstruction. More recently,  with a different approach, Miquel \textit{et al.}~\cite{miquel2003three} presented a study to show the usefulness of (3D) reconstruction of the heart from a series of two-dimensional (2D) MR images using a commercially available software. \\
Since 1985, several techniques have been developed to ameliorate the acquisition of Cine MRI for several medical applications, mainly by improving the temporal resolution and shortening the acquisition time while maintaining image quality (Waterton \textit{ et al.}~\cite{waterton1985magnetic}, Bluemke \textit{et al.}~\cite{bluemke1997segmented} Berggren \textit{et al.}~\cite{berggren2021super}). \\  
Cine MRI quality can be adversely affected by numerous imaging artifacts, resulting in missing or corrupted information, which will eventually lead to a decrease in the accuracy of the results. For instance, the acquisition of short axis view cine MR images requires multiple breath-holds.  Consequently, cardiac and respiratory motion, together with fast flowing blood, may cause inter-slice motion artifacts; also adjacent tissues with different properties or implants may cause local loss of signal in particular slices (\cite{ferreira2013cardiovascular, van2014cardiac}).
Several techniques have been proposed for the solution of this problem. For example, methods based on Generative Adversarial Networks (GANs) \cite{goodfellow2014generative} have contributed in solving various medical image problems, including the recovery of missing data. Lee \textit{et al.}~\cite{lee2019collagan} introduced an image imputation technique called Collaborative Generative Adversarial Network (CollaGAN). The method uses a single generator and discriminator network to approach the missing data using the remaining clean data set. They demonstrated their approach on several datasets including 280 brain images  and an additional T2 Fluid-Attenuated Inversion Recovery (FLAIR) sequence from 10 subjects. The CollaGAN  produced images of higher visual quality than the original dataset and showed the best performance compared to previous algorithms (see also~\cite{xia2021recovering}). \\
Another highly used technique is medical image interpolation. In particular, inter-slice interpolation, which is mainly used to reconstruct missing information between slices with as high accuracy as possible. Several approaches have already been proposed for medical image interpolation. Leng \textit{et al.}  \cite{leng2013medical} for example proposed a registration-based image interpolation technique. Their proposed method is divided into two steps: a registration process performed to construct a corresponding transformation between the slices, followed by interpolating intensity values along the matching lines to construct additional images between each two neighboring slices. They tested their method on different types of images, including two series of medical images: 1) a MR brain sequence, where they used their method to generate 3 additional interlayers between four subsequences of slices (with 40 slices per subsequence), 2) computed tomography (CT) chest sequence, where they removed all odd slices and reconstructed them by interpolation between the even slices (see also \cite{van2007nonrigid}).    
Horváth \textit{et al.}~\cite{horvath2017high} worked on approaching this problem with a more advanced technique, where they presented a high order slice interpolation method, which employs both object and intensity interpolation to interpolate an entire stack of images, solving the problem of combining higher order interpolations of structure motion and intensity. The proposed technique was tested on the human spinal cord along the neck acquired with a slice-wise inversion recovery MR sequence, where they used a leave-one-slice-out test to evaluate how well the left out slices can be interpolated.\\
Meantime, Ehrhardt \textit{et al.}~\cite{ehrhardt2006optical} derived an interpolation scheme from the optical flow equation mainly to generate images at predefined phases of the cardiac cycle in a cine MRI sequence of patients with myocardial infarction. The presented method consisted of two main steps: first, a nonlinear registration algorithm is used to determine the optical flow between the temporal images, then the calculated optical flow field is used to generate the interpolated images at the desired time. The obtained results were compared with previous techniques and it showed that the presented method outperformed several interpolation techniques.  Lin \textit{et al.}~\cite{lin2017slice} proposed a new algorithm for slice interpolation in MRI, called the decomposition-reconstruction (D-R) method. As its name indicates, the method first decomposes the information contained in the MRI and then reconstructs it based on the rule of organ consistency. The inter-slice interpolation algorithm was applied to a 3 neighboring MRI of the human throat and its performance was compared with three other interpolation techniques (\textit{linear} interpolation, \textit{spline} interpolation, and \textit{cubic} interpolation). The results showed that the D-R method was more suitable for MRI interpolation. Meanwhile Grevera \textit{et al.} \cite{grevera1998objective} made an objective comparison between 8 slice interpolation techniques using datasets from different medical applications, different body parts, different modalities and different patients. \\

In this work, the authors have taken a different path than the previous researchers, by introducing a novel technique combining the higher order dynamic mode decomposition (HODMD) with a singular value decomposition (SVD) based interpolation technique. The method is established on HODMD~\cite{le2017higher}, a tool often used in fluid dynamic applications that has been tested suitable by the authors~\cite{GROUN2022105384} to identify the main patterns driving the dynamics of the heart. More specifically, HODMD is introduced as a tool able to classify different heart pathologies in ecocardiography images based on variations of the heart rate and different patterns identified in the datasets analyzed. The HODMD, which uses a high order singular value decomposition (HOSVD)~\cite{tucker1966some} for data-dimensionality reduction and for cleaning the noise of the images, is paired with the interpolation technique to generate a tool with a threefold application: (i) to provide 3D reconstructions of the heart, (ii) to repair corrupted or missing information and finally, (iii) to develop a reduced order model (ROM), which is able to extend the original database that is formed by a small number of snapshots  and to generate new data, very useful to be used in future machine learning applications, which generally require an accurate and extensive database for the training.\\


The rest of the paper is organized as follows. S \ref{MandR}  introduces the dataset used in this work. In S \ref{Methodology}, the new algorithms for 3D reconstruction, data repairing and data modeling are introduced. The results and conclusions are presented in S \ref{Resul} and S \ref{Concl}, respectively.

\section{Acquisition of Cine MRI databases}\label{MandR}
 \begin{figure}[h!]
    \centering
    \includegraphics[width=7.5cm, height=4cm]{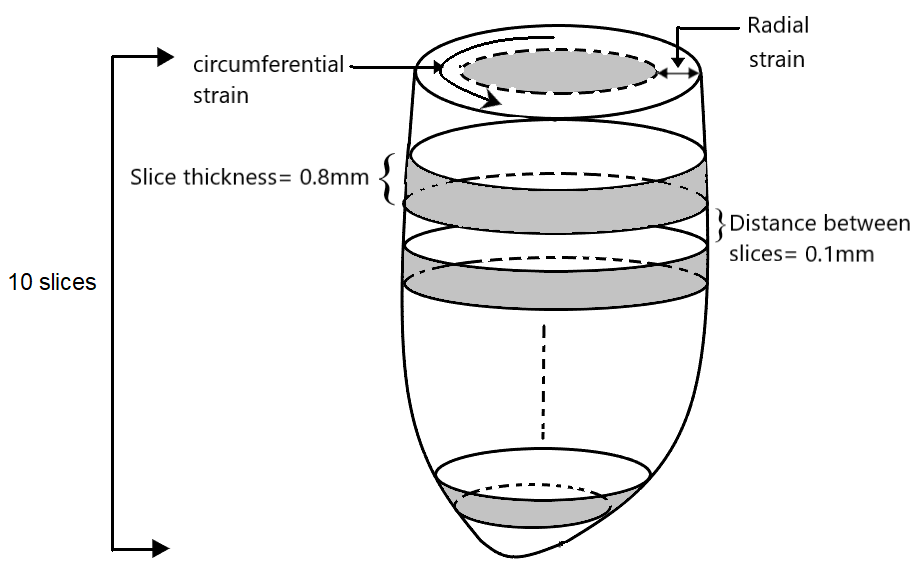}
    \caption{\textit{Sketch of a heart illustrating the acquisition of data using cine MRI.}}
    \label{Fig2}
  \end{figure}   

Hearts from healthy 10 months-old mice (and SFSR4 wild-type mice \cite{larrasa2021srsf4}), were used. All the data have been obtained from previous studies performed in accordance with protocols approved by the CNIC’s Institutional Animal Care and Research Advisory Committee of the Ethics Committee of the Regional Government of Madrid (PROEX177/17).
Mice were anesthetized by inhalation of 3-4\% isoflurane for induction and 1.5-2\% for maintenance), administered in 95\% oxygen using a nasal mask. Ophthalmic gel was placed in their eyes to prevent retinal drying. Normothermia was maintained with forced air warming the procedure. The core body temperature, electrocardiogram, heart rate and respiratory rate were continuously monitored with a CMR-compatible system for rodents.\\
In vivo cardiac images were acquired with a 7 by using 7-T Agilent/Varian scanner (Agilent, Santa Clara, CA, USA) equipped with a DD2 console, with an actively shielded 115/60 gradient. A surface coil was used for MR signal transmission and reception. MRI experiments were conducted by applying an \textit{ECG-triggered fast gradient echo cine} sequence with the following imaging parameters: TR/TE values around 125/1.25 msec (they are minimum values depending on heart rate and number of frames per heart cycle); field of view, 30 x 30 mm; acquisition matrix, 128x128; flip angle 15$^{\circ}$; 4 averages; 20 cardiac phases; 10 slices; slice thickness, 0.8 mm and gap, 0.1 mm.\\
Hence, 20 MR images (snapshots) of $128 \times 128 \times 1$ resolution were acquired for each slice, resulting a total number of 200 snapshot per dataset.

\section{Methodology}\label{Methodology}
This section presents the new methodology introduced for three-dimensional reconstruction, data repairing and reduced order modeling of MRI data.  The higher order dynamic mode decomposition will be briefly reviewed, whereas, the interpolation method will be extensively detailed. \\

For simplicity, the data can be are organized in matrix form in the following snapshots matrix :
\begin{equation}\label{Eq1}
\bm{ V}_1^K=[\bm{v}_1,\bm{v}_2,\dots ,\bm{v}_k,\dots ,\bm{v}_K], 
\end{equation}

where $\bm{v_k}$ is a snapshot collected at time $t_k$, with $ k = 1,\dots , K $. Hence, $\bm{ V}_1^K \in \mathbb{R}^{J \times K}$ , with $ J = N_x \times N_y$, where $  N_x \hspace{0.2cm} \textrm{and} \hspace{0.2cm} N_y$ are the total number of pixels on the $X$ and $Y$ directions, respectively. Generally, when dealing with such experimental data, it is common to have $K \ll J$. \\
In this work, the MR images of each slice were imported, reshaped into vectors and organized in an individual tensor, defined in discrete form as 
\begin{equation}\label{Eq2}
\resizebox{.9\hsize}{!}{$\bm{X}_{x_1,x_2,k}  \hspace{0.1cm} \textrm{for} \hspace{0.1cm}  x_1=1,\dots ,N_x  ;  x_2=1,\dots ,N_y \hspace{0.1cm} \textrm{and} \hspace{0.1cm}  k=1,\dots ,K,$} 
\end{equation}
where $ x_1 $ and $ x_2 $ represent the position of each pixel in the frame plane containing the image, for the horizontal and vertical components and $K$ is the number of snapshots. 
\subsection{Higher Order Dynamic Mode Decomposition}
Higher order dynamic mode decomposition (HODMD) is an extension of the well known method dynamic mode decomposition (DMD)~\cite{schmid2010dynamic}, which was first introduced the in the fluid dynamics community to identify flow patterns and characterize instabilities. HODMD \cite{le2017higher} was introduced for the analysis of complex data modeling non-linear dynamical systems \cite{le2020coherent}. The HODMD algorithm works on decomposing spatio-temporal data into a number of modes $\bm{u}_m$, which are weighted with the amplitudes $a_m$,  and oscillate in time $t$ with the frequencies $\omega_m$ and they may increase or decrease or stay persistent in time with the growth rate $\delta_m$, as follows:
\begin{equation} \label{Eq002}
\resizebox{.9\hsize}{!}{$ \bm{v}(\bm{x},t_k)\simeq \sum\limits_{m=1}^M a_m\bm{u}_me^{(\delta_m+i\omega_m)t_k} \hspace{0.1cm} \textrm{for} \hspace{0.1cm} k = 1, \dots , K ,$} 
\end{equation}
where $\bm{x}$ represents the spatial component of the database. In what follows, for the analysis of 2D images,  $x=(x_1,x_2)^T$, such that $x_1$ and $x_2$ are the horizontal and vertical coordinates of the image pixel (as mentioned before), respectively.

The HODMD algorithm can be summarized in two main steps (the algorithm is explained in detail in Ref. \cite{le2017higher}):\\

\textbf{Step 1- Dimentionality reduction:} this step is done by applying the singular value decomposition (SVD) to the snapshot matrix eq. (\ref{Eq1}): 
\begin{equation} \label{Eq3}
\bm{ V}_1^K \simeq \bm{W}\bm{\Sigma} \bm{T}^T ,
\end{equation}
such that, $\bm{W}$ and $\bm{T}$ are unitary matrices containing the spatial modes and the temporal modes, respectively. The matrix $\bm{\Sigma}$ contains in the diagonal the singular values $ \sigma_1,\dots ,\sigma_N$. The number of retained SVD modes $N$, is defined with respect to a certain tolerance $\varepsilon_{SVD}$ as 
\begin{equation} \label{Eq003}
\frac{\sigma_{N+1}}{\sigma_1} \leqslant \varepsilon_{SVD} .
\end{equation} 
Equation. (\ref{Eq3}) results the following \textit{reduced snapshot matrix} of dimension $N\times K$ : 
\begin{equation} \label{Eq4}
\bm{\widehat{V}}^K_1 = \bm{\Sigma} \bm{T}^T, \quad \textrm{with} \quad \bm{ V}^K_1 \simeq   \bm{W} \bm{\widehat{V}}_1^K . 
\end{equation}

\begin{figure}[h!]
    \centering
    \includegraphics[width=7cm, height=5cm]{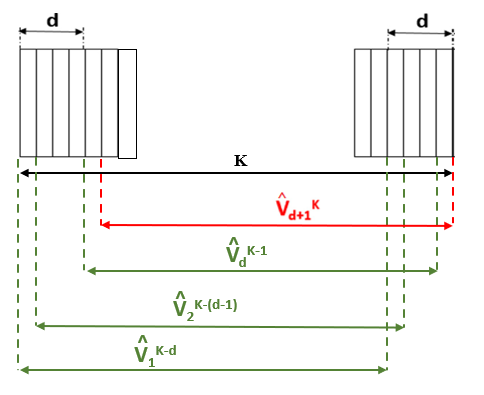}
    \caption{\textit{Reduced snapshot matrix} $\widehat{V}_1^{d+1}$ and the sliding window process defined in eq. (\ref{Eq5}).}
    \label{Fig1}
    
\end{figure}

\textbf{Step 2- The DMD-d algorithm:} This method can be seen as a combination of standard DMD and Takens' delay embedding theorem \cite{takens1981detecting}, where the standard DMD is extended using delayed snapshots.\\
In this step, \textit{ the higher order Koopman assumption} is applied to the reduced snapshot matrix:
\begin{equation} \label{Eq5}
 \bm{\widehat{V}}^K_{d+1}\simeq \bm{\widehat{R}}_1\bm{\widehat{V}}^{K-d}_1+\bm{\widehat{R}}_2\bm{\widehat{V}}^{K-d+1}_2 + \dots + \bm{\widehat{R}}_d\bm{\widehat{V}}^{K-1}_d ,  
\end{equation}
where $ \bm{ \widehat{R}}_k= \bm{W}^T \bm{R}_k \bm{W}  $ is \textit{the reduced Koopman operator}.\\
Equation (\ref{Eq5}) suggests that, as illustrated in Fig. (\ref{Fig1}), it is possible to compare DMD-d with the sliding window process,  usually carried out in power spectral density analysis, which allows DMD-d to calculate several temporal modes associated with a single spatial mode ameliorating the performance of DMD-d compared to the standard DMD. \\

The different \textit{Koopman operators} $ \bm{\widehat{R}}_1, \dots , \bm{\widehat{R}}_d$ in eq. (\ref{Eq5}) contain the dynamics of the system, therefore, all these operators are joined into a \textit{modified Koopman matrix} as :
\begin{equation}\label{Eq6}
\bm{\tilde{R}}\equiv\begin{bmatrix} 
\bm{0} & \bm{I} & \bm{0} & \dots & \bm{0} & \bm{0} \\
\bm{0} & \bm{0} & \bm{I} & \dots & \bm{0} & \bm{0} \\
\dots & \dots & \dots & \dots & \dots & \dots \\
\bm{0} & \bm{0} & \bm{0} & \dots & \bm{I} & \bm{0} \\
\bm{\widehat{R}}_1 & \bm{\widehat{R}}_2 & \bm{\widehat{R}}_3 & \dots & \bm{\widehat{R}}_{d-1} & \bm{\widehat{R}}_d    
\end{bmatrix}.
\end{equation}
The eigen-decomposition of the matrix eq. (\ref{Eq6}) is then used to obtain the frequencies $\omega_m$, growth rates $\delta_m$ and DMD modes $\bm{u}_m$. The amplitudes $a_m$ are calculated by least squares fitting, and the number of $M$ DMD  modes to retain is determined using the tolerance $\varepsilon_{DMD}$ as 
\begin{equation}\label{Eq006}
\frac{a_{M+1}}{a_1}  \leqslant \varepsilon_{DMD}  
\end{equation}    

\subsubsection{Handling noisy data}
Whenever the data being considered is very noisy (which is  precisely the  case here considered), it is worth to accomplish the dimensionality reduction using the more sophisticated higher order singular value (HOSVD) factorization.\\
As demonstrated in reference~\cite{le2017higher-1}, HOSVD offers better opportunities to clean the data. The only modification in the treatment so far substitutes the snapshot matrix (eq. (\ref{Eq1})) with the 
multidimensional snapshot tensor (eq. (\ref{Eq2})).\\
The HOSVD algorithm applies SVD along each spatial direction of the \textit{tensor}, which yields to the following decomposition:
\begin{equation}\label{Eq7}
  \bm{X}_{x_1x_2k}\simeq \sum\limits_{p1=1}^{P_1}\sum\limits_{p_2=1}^{P_2}\sum\limits_{n=1}^{N} \bm{S}_{p_1p_2n}\bm{W}^{(1)}_{x_1p_1}\bm{W}^{(2)}_{x_2p_2}\bm{\mathsf{T}}_{kn},  
\end{equation}
where $ \bm{S} $ is the \textit{core tensor} and the columns of $\bm{W}^{(1)}, \bm{W}^{(2)} $ are called spatial modes (connected to the number of component, streamwise and normal direction) and columns of $\bm{\mathsf{T}}_{kn}$ are the temporal modes of the decomposition. The process of obtaining these three sets of modes (by applying SVD to the three matrices whose columns are the associated fibers of the tensor), results three sets of non-zero singular values $\sigma_{p_1}^{(1)} , \sigma_{p_2}^{(2)} \hspace{0.2cm} \textrm{and} \hspace{0.2cm} \sigma_{n}^{(t)} $. \\ 
Hence, eq. (\ref{Eq7}) can be rewritten as 
\begin{equation}\label{Eq8}
  \bm{X}_{x_1x_2k}\simeq \sum\limits_{n=1}^{N} \bm{\widehat{S}}_{x_1x_2n}\bm{\widehat{\mathsf{T}}}_{kn},  
\end{equation}
where the spatial modes $\bm{\widehat{S}}_{x_1x_2n}$ and the rescaled temporal modes $\bm{\widehat{ \mathsf{T}}}_{kn}$ are defined as
 \begin{equation}\label{Eq9}
\resizebox{.89\hsize}{!}{$ \bm{\widehat{S}}_{x_1x_2n} =  \sum\limits_{p1=1}^{P_1}\sum\limits_{p_2=1}^{P_2} \bm{S}_{p_1p_2n}\bm{W}^{(1)}_{x_1p_1}\bm{W}^{(2)}_{x_2p_2} / \sigma_n^t \hspace{0.2cm} \textrm{and} \hspace{0.2cm} \bm{\widehat{ \mathsf{T}}}_{kn} = \sigma_n^t \bm{\mathsf{T}}_{kn} $}
\end{equation}
 We determine the number of modes to retain $N$ from each one of the spatial modes, using the tolerance $\varepsilon_{SVD}$ as in eq. (\ref{Eq003}). Finally, step 2 of the standard HODMD is applied to the  temporal modes $ \bm{\mathsf{T}} $. 
The method used in this previous case is called the \textit{multidimensional iterative HODMD}. The iterative algorithm is mainly applying multi-dimensional HODMD iteratively over the reconstructed data (eq. (\ref{Eq002})) until the number of modes is fixed between two successive iterations. For more details see Ref.~\cite{le2017higher-1}. 
\subsection{Interpolation Technique}\label{Interp_SVD}
 \begin{figure*}[h!]
    \centering
    \includegraphics[width=18cm, height=9cm]{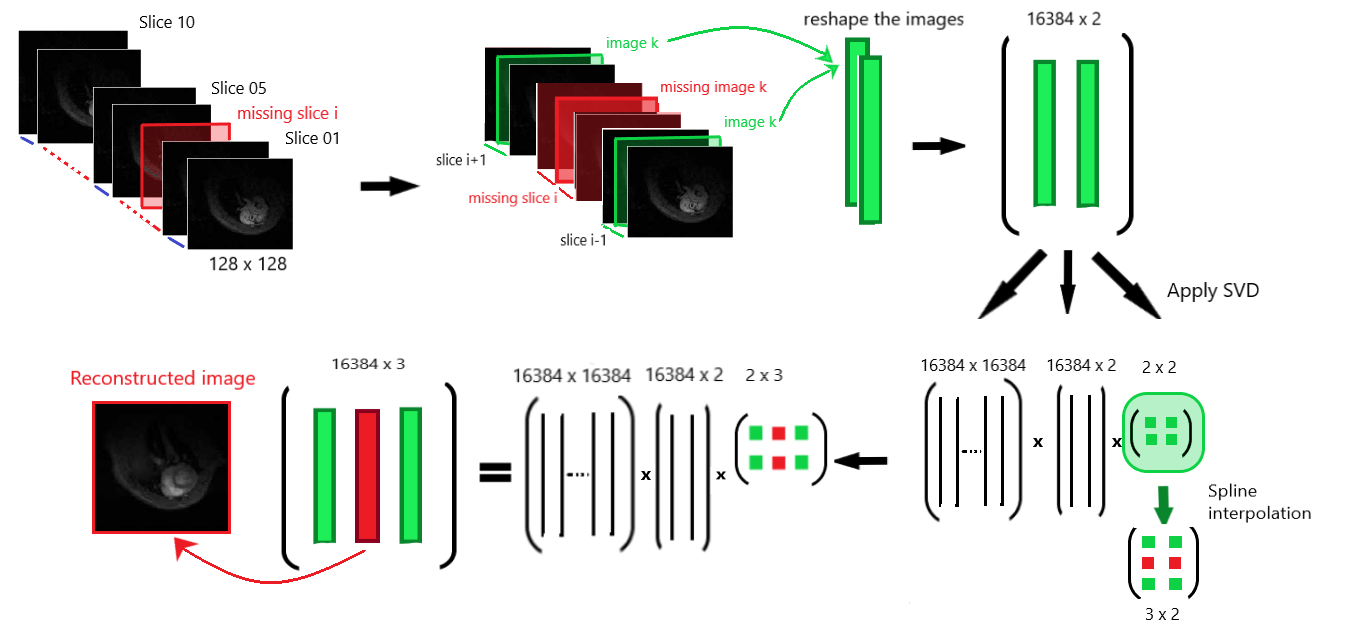}
    \caption{Schematic diagram for the interpolation technique pipeline.
The k-\textit{th} snapshot (frame) from corrupted/missing slice \textit{i} can be reconstructed by the following procedure:
First, the k-\textit{th} frames from the neighboring $i-1$ and $i+1$ slices are reshaped into vectors, which are arranged into the tall and skinny matrix $\bm{\mathit{M}}$.
Second, this matrix is then factorized using SVD as  $\bm{\mathit{M}}= \bm{W}\bm{\Sigma} \bm{T}^T$.
Third, \textit{spline} interpolation is performed on the matrix of the right singular vectors, providing the matrix $\tilde{T}$.
Finally, the reconstructed data is  retrieved as $\bm{\tilde{\mathit{M}}}= \bm{W}\bm{\Sigma} \bm{\tilde{T}}^T$. }
    \label{Fig001}
  \end{figure*}
The interpolation technique used in our work was inspired by an image resolution enhancement method. The method was developed by R{\"o}vid \textit{et al.}~\cite{rovid2011hosvd}, where they introduced a new data representation domain related to HOSVD \cite{tucker1966some}.\\
In the following, we explain how we applied this concept on a two dimensional array, using SVD \cite{golub1971singular} instead of HOSVD for the reconstruction of corrupted or missing images.\\ 
 Figure \ref{Fig001} presents a sketch
describing the interpolation technique applied to reconstruct the
missing MR images. Details about the technique are explained next.\\
 
To reconstruct the missing information of an image $f_k$ (for $k = 1, ..., K$, with $K$ is the number of snapshots) of a slice $i$, we consider  a $J\times 2$ matrix denoted $\bm{\mathit{M}}$. The matrix contains in columns the pixels of the $k$\textit{th} image of slices $i-1$ and $i+1$, respectively, with $J = N_x\times N_y$. First, we apply the SVD the matrix $\bm{\mathit{M}}$ as follows:
\begin{equation}\label{Eq17}
\bm{\mathit{M}}= \bm{W}\bm{\Sigma} \bm{T}^T ,
\end{equation}
where $\bm{W}$ $\in \bm{\mathbb{C}} ^{J\times J}$ and $\bm{T}$ $\bm{\in \mathbb{C}}^{2\times 2}$ are unitary matrices and $J$ is number of SVD modes. The matrix $\bm{\Sigma \in \mathbb{R}} ^{J\times 2} $ contains real, non negative singular values on the diagonal and zeros off the diagonal.\\
To generate the missing image, the matrix $\bm{T}$ will be updated, where the number of columns will remain the same and the number of lines will be extended to 3. Lets note $\bm{\tilde{T}}$ the updated matrix, the lines $\bm{\tilde{T}}_s $ with  $1\leq s \leq 3$, will be determined as follows:  $\bm{\tilde{T}}_1 = \bm{T}_1 $,  $\bm{\tilde{T}}_3 = \bm{T}_2 $, meanwhile, \textit{spline} interpolation is used to determine the elements of $\bm{\tilde{T}}_2$. When all the lines are determined, eq. (\ref{Eq17}) is used to obtain a new matrix containing 3 columns, representing the pixels of the two previous images plus the pixels of the  reconstructed missing image.\\
This procedure is repeated for each time step $t_k$ in order to reconstruct all the snapshots of the missing slice. \\

To validate the method, some slices are removed from the database, and the interpolation technique introduced is used to reconstruct such missing slices. The reconstructed slice is compared with the original slice using the \textit{relative root mean square} error (RRMSE), which is defined as follows:
\begin{equation}\label{eq_err}
 RRMSE= \frac{•\| \bm{Y}^{approx}-\bm{Y}\|_2}{\|\bm{Y}\|_2} ,
\end{equation}

where $ \bm{Y}^{approx}$ is the new reconstructed i\textit{th} slice, $\bm{Y}$ is the same i\textit{th} slice in the original database and $\|.\|_2$ is the Frobenius norm for matrices and tensors. 
\subsection{HODMD as a reduced order model}
This section explains how the HODMD was applied over the MR images, providing a 3D reconstruction of the heart, varying in time using the DMD expansion eq. (\ref{Eq002}).
\\
\begin{itemize}
\item First, each slice is arranged in tensor form as represented in eq. (\ref{Eq2}). The HODMD was applied to each tensor, using in process HOSVD to clean the data, by applying SVD to every spatial dimension with respect to the tolerance (eq. (\ref{Eq003})). The HODMD will result the reconstructed slices, which will be joined in one tensor, presented as follows
\begin{equation}\label{Eq0001}
\resizebox{.9\hsize}{!}{$\bm{X}_{x_1,x_2,i,k}  \hspace{0.1cm} \textrm{for}\hspace{0.1cm}  x_1=1,\dots ,N_x  ;  x_2=1,\dots ,N_y  ;  i=1,\dots ,I \hspace{0.1cm} \textrm{and}\hspace{0.1cm}  k=1,\dots ,K ,$}
\end{equation}
where $ x_1 $ and $ x_2 $ represent the position of each pixel in the plane containing the image (as defined before), $I$ is the number of slices and $k$ correspond to the snapshot number.
\item Next, HODMD was applied one more time to the tensor containing all the reconstructed slices, with a change in the temporal term, where the expansion defined in eq. (\ref{Eq002}) will not be only applied to its temporal term $[t_1,t_K]$, but to a larger one $[t_{K+1}, t_{R}]$, with $R > K$, exceeding the original number of time steps. In this way, it is possible to obtain a new database, which represents the evolution of the 3D reconstruction, with extra reconstructed snapshots covering more than one cardiac cycle (see more details about this methodology and possible applications in Ref.(\cite{le2017higher-2, le2019prediction})).\\ 
\end{itemize}

\section{Results}\label{Resul}
\subsection{3D reconstruction}
\begin{figure}[h!]
    \centering
    \includegraphics[width=9cm, height=6.5cm]{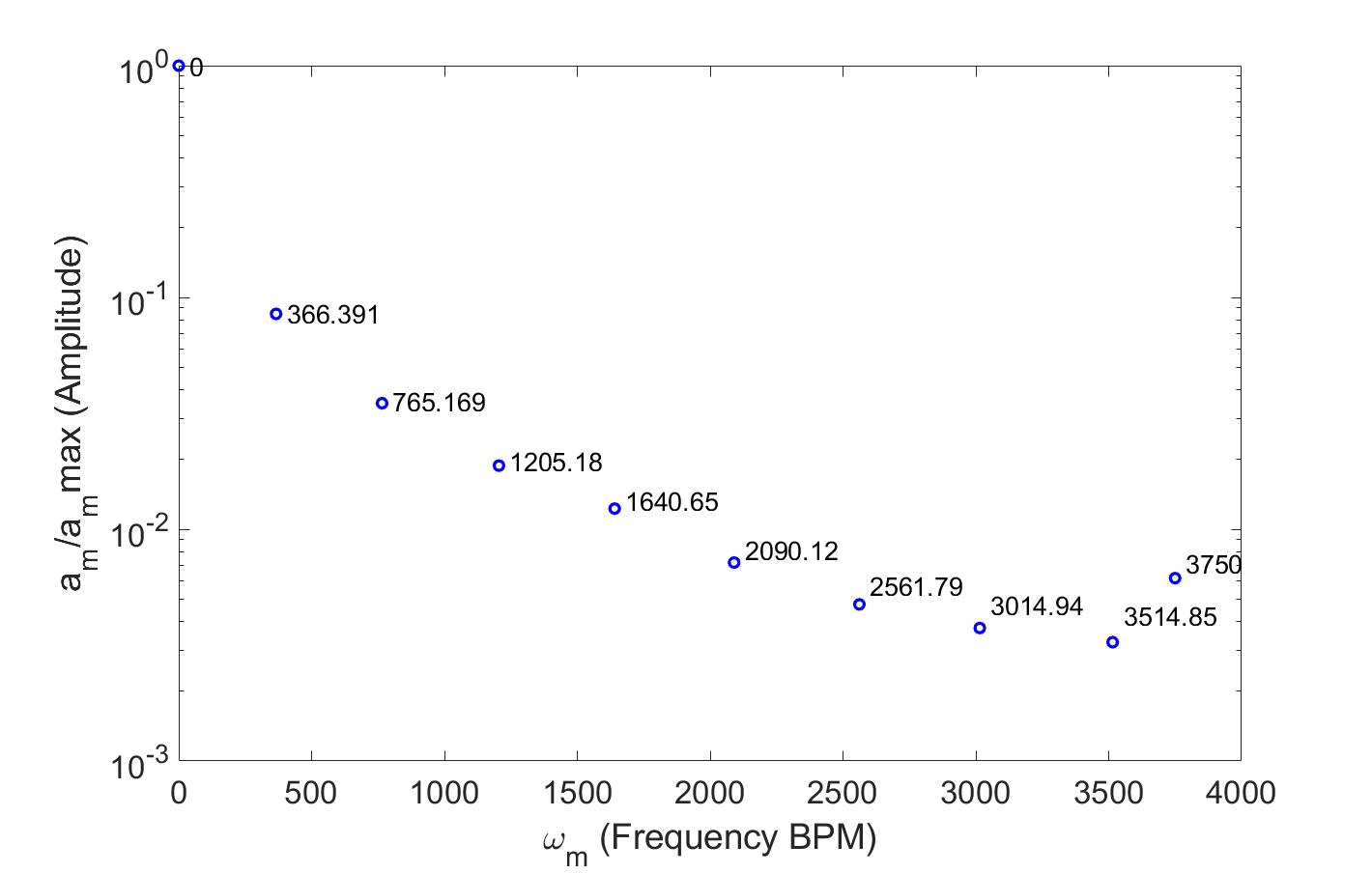}
    \caption{\textit{Frequencies identified by HODMD applied to the reconstructed MRI tensor.}}
    \label{Fig3} 
\end{figure}

For the 3D reconstruction, all the slices went through the same procedure. Each slice was imported, then all the snapshots were extracted, reshaped and arranged in a fourth-dimensional tensor as in eq. (\ref{Eq0001}). All the tensors were analyzed using the HODMD algorithm with the following parameters: (i) the number of snapshots $K=20$ representing a complete heartbeat, as mentioned before,  (ii) the tolerances $\varepsilon_{SVD}=\varepsilon_{DMD}= 5\times 10^{-4} $ , which are related to the amplitude truncation and the noise level, respectively (these tolerances have been selected based on our previous work  \cite{GROUN2022105384}, where we have already used the HODMD algorithm to analyze medical images), (iii) the time step $\Delta t$ is estimated to be $\Delta t=8\times 10 ^{-3}$ and (iv) the index $d$ is set to $d= 2$ for all the slices (following the calibration tips presented in Ref. \cite{vega2020higher}). 
HODMD was applied first to each one of the slices separately, and then to all the slices simultaneously (in tensor form, as in eq. (\ref{Eq2})). As expected, the solution is periodic (as shown in Fig. (\ref{Fig3})). In both cases, the method was able to capture the leading frequency (the frequency of the heart) as ~350-370 beat per minute (BPM). The differences found between the different calculations are connected to the presence of noise. It is remarkable that, the number of snapshots used for this analysis was only 20, representing a complete heartbeat. Using larger number of snapshots will imply obtaining more accurate results, less noisy, hence, the value of the leading frequency will be similar in all the tests carried out.
\FloatBarrier
 
\begin{figure}[h!]
    \centering
    \includegraphics[width=7cm, height=5.7cm]{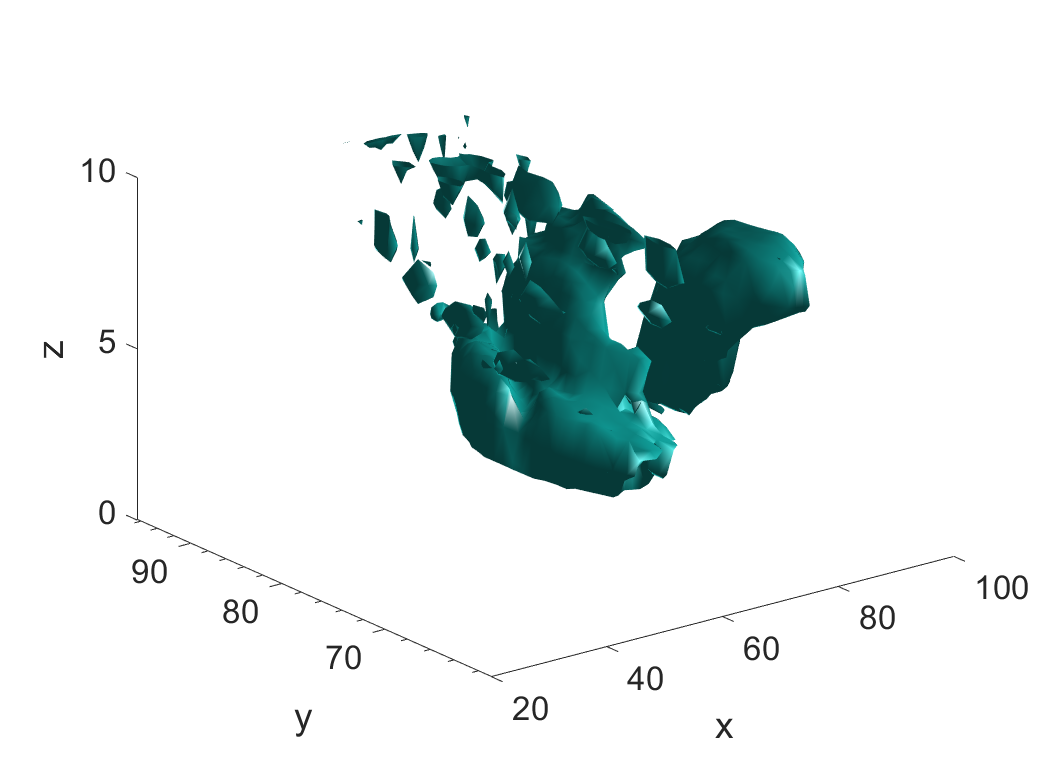}
    \caption{\textit{3D reconstruction of the healthy heart.}}
    \label{Fig4}
\end{figure}

HODMD (and HOSVD) combined with the interpolation algorithm provides the 3D reconstruction of the heart, as presented in Fig. (\ref{Fig4}). The method identifies the leading frequency of these 3D data as $\sim 370$ BPM, in good agreement with the results obtained when the algorithm was applied in the original core database. The 3D visualization shown in Fig. (\ref{Fig4}) is carried out using the MATLAB\textsuperscript{\textregistered}~\cite{MATLAB:R2020b} visualization tool "\textit{isosurface}". 
Finally, HODMD is used as a ROM to generate new data describing different heart cycles. More specifically, from a small database formed by 20 snapshots, 80 new snapshots have been generated (hence the total number of snapshots of the database is now 100). Figure (\ref{Fig5}) shows 12 different snapshots of the heart cycle.  (Video available in supplementary material).

\begin{figure}[h!]
	\centering
	\textbf{ \scriptsize Original snapshot 01 \hspace{0.01cm} \scriptsize New snapshot 25 \hspace{0.01cm} \scriptsize New snapshot 65 }\\
\rotatebox{90}{\hspace{0.8cm}\textbf{Slice 01}}\hspace{0.0001cm}
\subfloat[  ]
{\includegraphics[width=2.7cm, height=2.7cm]{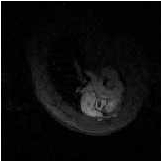}} \hspace{0.0001cm} 
\subfloat[ ]{\includegraphics[width=2.7cm, height=2.7cm]{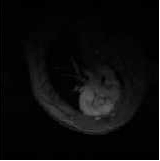}}\hspace{0.001cm} 
\subfloat[  ]{\includegraphics[width=2.7cm, height=2.7cm]{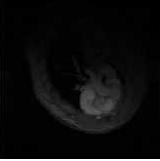}}\\
\rotatebox{90}{\hspace{0.8cm}\textbf{Slice 03}}\hspace{0.1cm}
\subfloat[  ]
{\includegraphics[width=2.7cm, height=2.7cm]{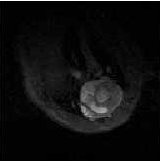}} \hspace{0.01cm} 
\subfloat[ ]{\includegraphics[width=2.7cm, height=2.7cm]{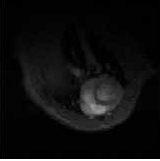}}\hspace{0.01cm}
\subfloat[ ]{\includegraphics[width=2.7cm, height=2.7cm]{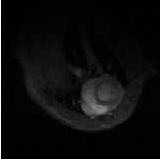}}\\
\rotatebox{90}{\hspace{0.8cm}\textbf{Slice 07}}\hspace{0.1cm}
\subfloat[ ]
{\includegraphics[width=2.7cm, height=2.7cm]{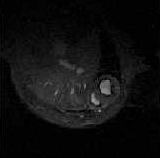}} \hspace{0.01cm} 
\subfloat[ ]{\includegraphics[width=2.7cm, height=2.7cm]{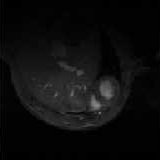}}\hspace{0.01cm}
\subfloat[ ]{\includegraphics[width=2.7cm, height=2.7cm]{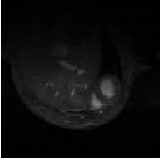}}\\
\rotatebox{90}{\hspace{0.8cm}\textbf{Slice 10}}\hspace{0.1cm}
\subfloat[ ]
{\includegraphics[width=2.7cm, height=2.7cm]{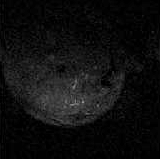}} \hspace{0.01cm}
\subfloat[ ]{\includegraphics[width=2.7cm, height=2.7cm]{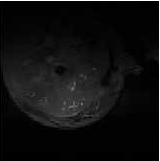}}\hspace{0.01cm}
\subfloat[ ]{\includegraphics[width=2.7cm, height=2.7cm]{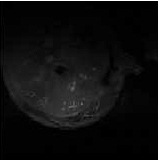}}\\

\centering
	\caption{Different reconstructed snapshots using a ROM based on HODMD. The heartbeat cycle is defined by 20 snapshots. }	
	\label{Fig5}
\end{figure}
\subsection{Data repairing using HODMD}

\begin{figure}[h!]

    \centering
    \textbf{ Original 7\textit{th} slice \hspace{1cm} Interpolated 7\textit{th} slice }\\
\subfloat[ ]{\includegraphics[width=3.3cm, height=3.3cm]{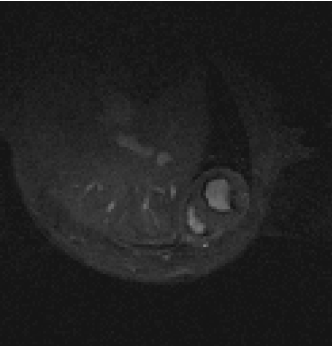}} \hspace{0.3cm} 
\subfloat[ \label{Fig7.a}]{\includegraphics[width=3.3cm, height=3.3cm]{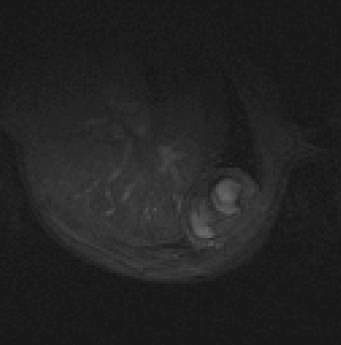}} \\	
\subfloat[ ]{\includegraphics[width=3.3cm, height=3.3cm]{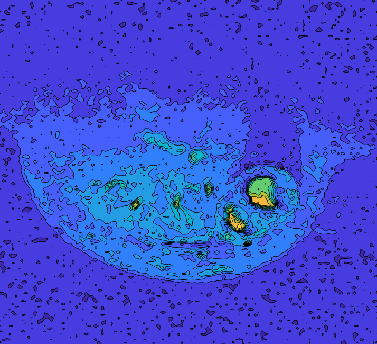}}\hspace{0.3cm}
\subfloat[ \label{Fig7.b}]{\includegraphics[width=3.3cm, height=3.3cm]{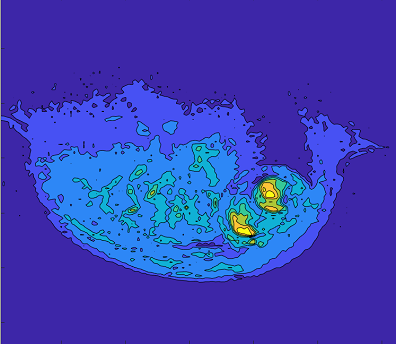}}\\
\caption{A comparison between snapshot (01) taken from original slice 07 and  interpolated slice 07. (a) and (c) correspond to the original snapshot, and (b) and (d) to the reconstruction. The intensity of the image is presented in (c) and (d). The colors correspond to 1 (yellow), 0.5 (green) and 0 (blue).}
    \label{Fig7} 

\end{figure}

\begin{figure}[h!]
	\centering
	\subfloat[\label{Fig6.a}]{\includegraphics[width=7.5cm, height=5.5cm]{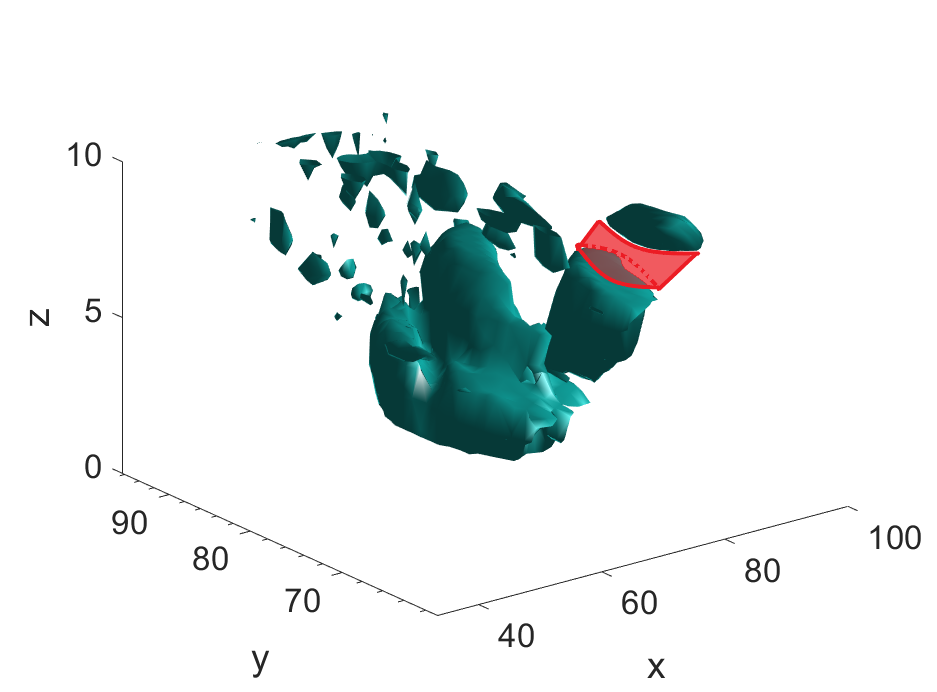}} \\
\subfloat[ \label{Fig6.b}]{\includegraphics[width=7.5cm, height=5.5cm]{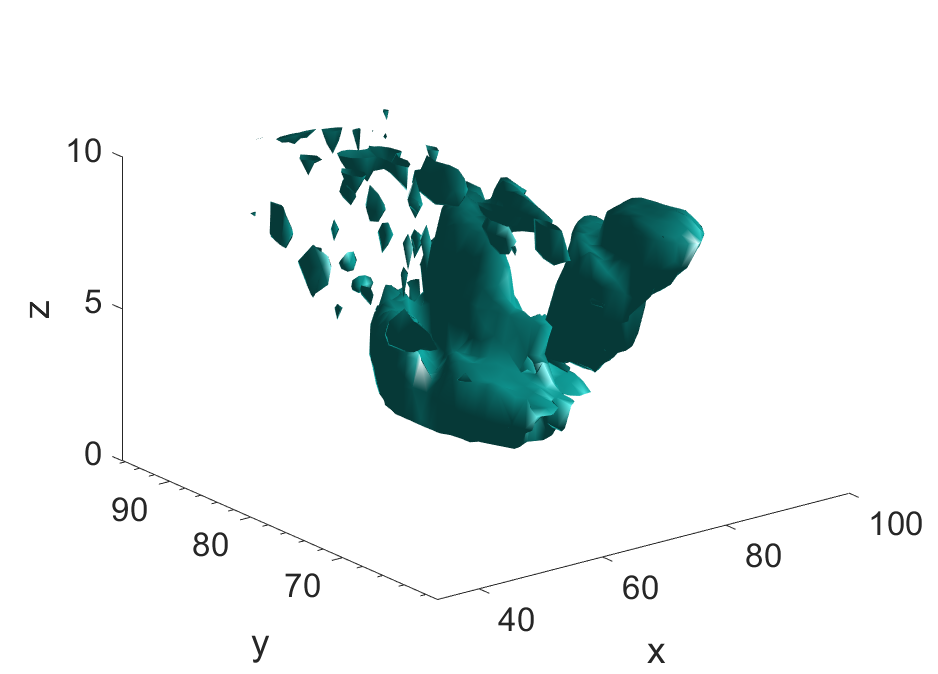}}\\
\centering
	\caption{3D reconstruction of the heart before and after the missing information recovery step. (a) is 3D visualization of the heart after removing the information of the 7\textit{th} slice marked in red. (b) is 3D visualization of the heart with the interpolated 7\textit{th} slice. }	
	\label{Fig6}
\end{figure}

In this section, the performance of the interpolated techniques combined with HODMD is shown on the tensor containing all the reconstructed slices, The main goal is to show the capabilities of HODMD as a data repairing method.\\
The original database is formed by 10 slices. To test the performance of the method, slice 7 was removed. Then, using the interpolation techniques introduced in section \ref{Interp_SVD}, these images are reconstructed. Fig. \ref{Fig7} compares the original database with the reconstruction. As seen the two images present a qualitative similar shape and intensity. The noise is  the main difference found between the two slices, hence the reconstructed image is clean.  As a final step, the 3D reconstruction of the heart is carried out, considering the reconstruction of the new slice as presented in Fig. (\ref{Fig6}). The RRMSE comparing the 3D reconstructions using the original data and the data with the 7th slice interpolated is 0.09. Beside the interpolation technique used in this work (\textit{spline}), 08 other interpolation techniques were tested (\textit{kriging}, \textit{linear},  \textit{cubic}, \textit{pchip},  \textit{makima}, \textit{nearest}, \textit{next} and \textit{previous}). The techniques: \textit{kriging},  \textit{linear},  \textit{cubic}, \textit{pchip} and \textit{makima} delivered similar reconstructions to the ones presented in this section, with the same reconstruction error. Meanwhile, \textit{nearest}, \textit{next}, \textit{previous} showed poor performance, increasing the error to 0.12. Hence, we should explore using other type of interpolation techniques  in future works to try to reduce this error, but this remain as an open topic. (The full videos showing the 3D reconstructions are available in supplementary material).
\FloatBarrier

Similar test have been carried out in slices 02, 05 and 09 (not shown for the sake of brevity), where the RRMSE for the 3D  reconstruction was always smaller than 0.09, showing the robustness of this methodology.
 \section{Robustness of the HODMD method}
\begin{figure}[h!]
	\centering
	\subfloat[ ]{\includegraphics[width=6cm, height=5.5cm]{3D_Recon}} \\
\subfloat[ ]{\includegraphics[width=6cm, height=5.5cm]{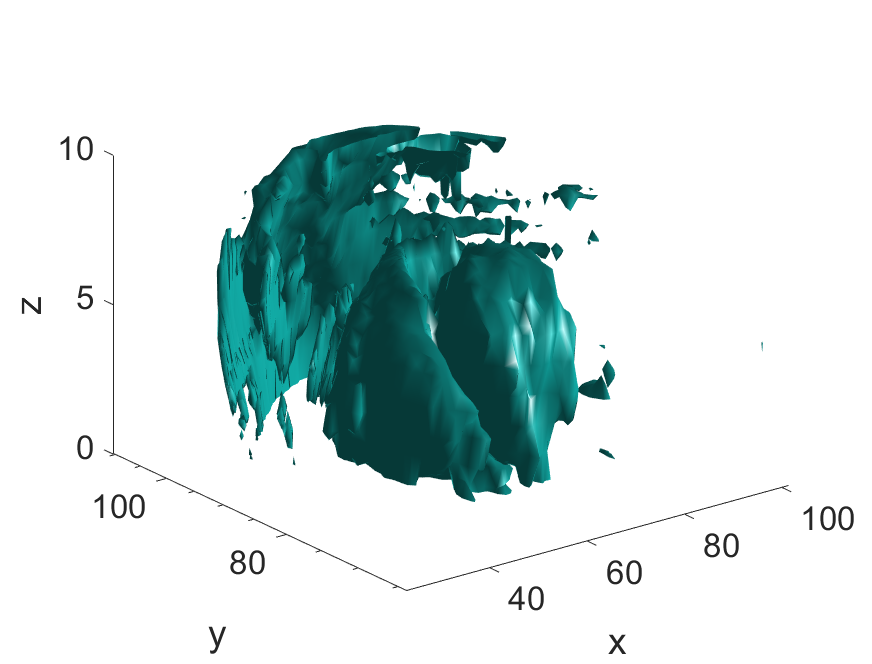}}\quad
\subfloat[ ]{\includegraphics[width=6cm, height=5.5cm]{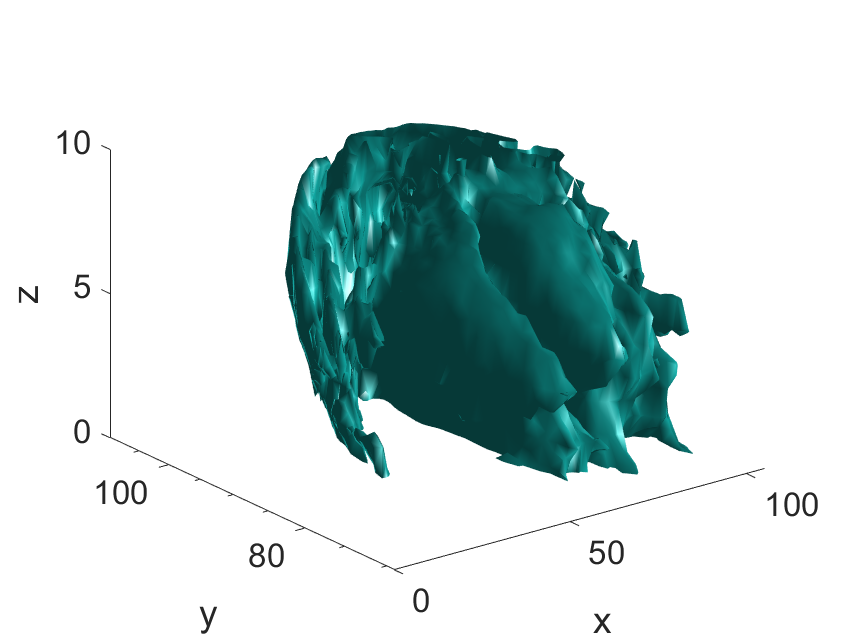}} \\

\centering
	\caption{A comparison between 3D layouts of the healthy and hypertrophic hearts. (a) 3D layout of the healthy heart, (b) 3D layout of the first hypertrophic heart and (c) 3D layout of the second hypertrophic heart.  }	
	\label{Fig8}
\end{figure}

To test the robustness of the method introduced, HODMD is applied over two more datasets. The two extra datasets were taken from mice afflicted with a cardiac disease, which is hypertrophy. Similarly to the first data set, the HODMD was used to analyze each slice of the new databases in order to obtain the reconstructed slices, which are used for the 3D layout of the afflicted hearts. In this case the 3D reconstruction allowed us to notice the difference between the sizes of the healthy and unhealthy hearts. As seen in Fig. (\ref{Fig8}) we can notice the enlargement of the hearts with hypertrophy when compared with the regular thin shape of the healthy one, as well as the stiffness of the hypertrophic hearts compared with the smooth and regular contraction and relaxation movement of the healthy heart (The videos are available in supplementary material ). The connection of this stiffness and its relationship with different pathologies remains as open topic to study in future works.
\FloatBarrier

 \section{Conclusion:}\label{Concl}
 In this work, the higher order dynamic mode decomposition (HODMD) has been combined with an efficient interpolation tool to explore the heart dynamics and build a 3D reconstruction of the heart. Sets of MR images taken from 10 different slices of the heart were analyzed first by the HODMD algorithm. Despite the limited amount of data (20 snapshots per slice), HODMD was able to capture the heart rate for each slice. This methodology, widespread in the field of fluid dynamics, 
has been leveraged for the first time (to the best of the authors' knowledge) for data repairing and 3D reconstruction of
medical imaging. Additionally, the methodology has allowed to develop a reduced order model of the heart. The interpolation algorithm introduced was used to reconstruct corrupted images and recover missing information from several  different slices. The algorithm presented is capable of satisfactorily reconstruct a completely missing slice from the information of neighboring slices. The technique was tested on 4 different slices, giving similar results with an error never exceeding 9\% for the full 3D reconstruction, proving the robustness of the method. These error values are mainly connected with the noise calculated or the original image compared to the reconstructed image, which is completely free of noise. To develop a reduced order model (ROM), we start from an original database, containing only 20 snapshot per slice, and by changing the temporal term in the HODMD algorithm, it was possible to generate a new database consisting of 100 snapshot per slice. The additional generated snapshots were used to extend the evolution of the 3D reconstruction of the heart. The robustness of this method was tested on two other, different  datasets, delivering the same positive results and proving the utility of the novel technique to identify different patterns in healthy and unhealthy hearts. 
\section{Acknowledgment:}
This work has been supported by  SIMOPAIR
(Project No. REF: RTI2018-097075-B-I00) funded by
MCIN/AEI/10.13039/501100011033
and by  the European Union's Horizon
2020 research and innovation program under the Marie Skłodowska-Curie
Agreement number 101019137— FLOWCID.\\
S.L.C. acknowledges the grant PID2020-114173RB-I00 funded by MCIN/AEI/10.13039/501100011033.\\
"Biomedical Imaging has been conducted at the Advanced Imaging Unit of the CNIC (Centro Nacional de Investigaciones Cardiovasculares Carlos III), Madrid, Spain." "This project used the ReDIB ICTS infrastructure TRIMA@CNIC, Ministerio de Ciencia e Innovación (MCIN)."


\end{document}